\documentclass[journal,10pt]{IEEEtran}

\usepackage{cite}
\usepackage{flushend}

\usepackage{amsmath,amssymb,amsfonts,bm,subfigure,lipsum}
\usepackage{stfloats}
\usepackage{algorithm}
\usepackage{algorithmic}
\usepackage{mathrsfs}
\usepackage{graphicx}
\usepackage[table]{xcolor}
\usepackage{textcomp}
\usepackage{xcolor}
\usepackage{epstopdf}
\usepackage{color}
\usepackage{booktabs}
\usepackage{amsmath}
\usepackage{bbm}
\usepackage{multicol}
\usepackage{multirow}
\usepackage[english]{babel}
\usepackage{upgreek}
\usepackage{float}
\usepackage{colortbl}
\usepackage[table]{xcolor}
\usepackage{cuted}

\def\BibTeX{{\rm B\kern-.05em{\sc i\kern-.025em b}\kern-.08em
    T\kern-.1667em\lower.7ex\hbox{E}\kern-.125emX}}
\usepackage{graphicx,graphics,color,epsfig,subfigure,graphpap,rotate}
\usepackage{times, verbatim, subfigure, epsfig, graphicx, latexsym}
\usepackage{url}
\usepackage{subfigure}

\begin{document}
\title{Rethinking: Deep-learning-based  Demodulation and Decoding}

\author{Boxiang~He, Zitao Wu, and Fanggang~Wang

}%

\maketitle

\begin{abstract}
Over the past few decades, the information theory community has worked to develop modulation and encoding that achieve the Shannon capacity in the constraint of the low implementation complexity. In this paper, we focus on the demodulation/decoding of the complex modulations/codes that approach the Shannon capacity. Theoretically, the maximum likelihood (ML)  algorithm can achieve the optimal error performance whereas it has $\mathcal{O}(2^k)$  demodulation/decoding complexity with $k$ denoting the number of information bits. Recent progress  in deep learning provides a new direction to tackle the demodulation and the decoding. The purpose of this paper is to analyze the feasibility of
the neural network  to demodulate/decode the complex modulations/codes close to the Shannon capacity and characterize  the error performance and the complexity of the neural network. Regarding the neural network demodulator,  we use the golden angle modulation (GAM), a promising modulation format that can offer the Shannon capacity approaching performance, to evaluate the demodulator. It is observed that  the neural network demodulator can get a close performance to the ML-based method while it suffers from the lower complexity order in the low-order GAM. Regarding the neural network decoder, we use  the Gaussian codebook, achieving the Shannon capacity, to evaluate the decoder. We also  observe that the neural network decoder achieves the error performance close to the ML decoder with a much lower complexity order in the small Gaussian codebook. Limited by the current training resources, we cannot evaluate the performance of the high-order modulation and the long codeword. But, based on the results of the low-order GAM and the small Gaussian codebook, we boldly give our conjecture: the neural network demodulator/decoder is a strong candidate approach for demodulating/decoding the complex modulations/codes close to the Shannon capacity owing to the error performance of  the near-ML algorithm and the lower complexity.

\end{abstract}
\section{Introduction}
\IEEEPARstart{I}{n} 1948, Shannon has provided the  mathematical theory of communication\cite{b1}. Guided by the theory of Shannon, the researchers have tried their best to develop  communication techniques that meet the Shannon limit in the following decades.

Various modulation formats have been developed and applied, such as pulse amplitude modulation (PAM), phase shift keying (PSK), quadrature amplitude modulation (QAM), amplitude-PSK (APSK), star-QAM, etc. Among them, QAM is the most well-known modulation scheme and is deployed in various communication systems. However, there is  an asymptotic shaping-loss of $\pi{\rm e}/6$ ($\approx1.53$ dB) in additive Gaussian noise channel according to Shannon-Hartley theorem \cite{b14}. To overcome the shaping-loss, probabilistic-shaping and geometric-shaping  techniques are studied to design novel modulation schemes. They aim at optimizing the location and probability of occurrence of constellation points to achieve complex Gaussian distribution. Examples of some  early works are trellis shaping \cite{b15}, nonuniform-QAM \cite{b16}, and asymmetric constellations\cite{b17}. The recent work  in \cite{b18} proposes the golden angle modulation (GAM) that asymptotically approaches the
Shannon capacity as the number of signal constellation points
grows.  Meanwhile, the biggest  advance in coding theory is the discovery of low-density parity-check (LDPC) codes\cite{b2}, turbo codes\cite{b3} and polar codes\cite{b4}, where the  polar codes asymptotically achieve the Shannon limit.

Regarding the demodulation and the decoding, the maximum likelihood (ML) algorithm is universal and optimal for the demodulation and the decoding in theory. However, the ML-based demodulation/decoding is impractical with the exponential complexity. Thus, it is attractive for researchers to design a new demodulation/decoding algorithm that approaches the  error performance of the ML algorithm with the lower complexity.


In demodulation, a generalized bit level demodulation scheme for M-ary QAM systems is proposed in \cite{b19}, which significantly reduces  the complexity and has almost the same bit error rate (BER) performance as the ML algorithm.  The authors in \cite{b20} propose  the deep convolutional neural network to demodulate the  Rayleigh-faded signal and the results show the deep convolutional neural network has a lower bit error
probability compared to other demodulators such as the support vector machine. In \cite{b21}, the authors propose the deep-learning-based demodulator in short-range multipath channels, where  the deep belief network and the stacked autoencoder are applied to their demodulation system. In \cite{b22}, the mixed neural network using the convolutional neural network  and recurrent neural network is proposed to demodulate the received signal. The authors in \cite{b23} propose a demodulator based on the convolutional neural network with variable input and output length. The  hard bit information is used to train the convolutional neural network and the log probability ratio based on the output layer of the trained network is proposed to realize the soft demodulation.

In decoding,   Gallager proposes the belief propagation for the LDPC code decoding, which is an iterative soft-decoding algorithm \cite{b5}. In \cite{b6}, the authors propose an  iterative turbo decoder including two soft-input soft-output decoders. In \cite{b4}, the successive cancellation algorithm is designed for the polar code decoding. In addition to the above decoding algorithms for specific codes, the universal decoding algorithms also have been widely studied. In \cite{b7}, the authors propose a universal and near-ML decoder based on the reordering of the received symbols according to their reliability measure. However, the decoder involves the  Gauss-Jordan elimination and thus suffers from the high computation complexity. In \cite{b8}, the authors introduce a universal decoder that rank-orders noise sequences from most likely to least likely. Further, the decoder in \cite{b8} can realize the ML decoding for arbitrary codebooks in discrete channels with or without memory.
Owing to the powerful fitting ability of the deep neural network, it shows a good performance on the decoding. The authors in \cite{b9} propose the deep learning method for improving the belief propagation (BP) algorithm  by assigning weights to the edges of the Tanner graph, which can achieve an error performance close to the traditional  BP decoders with fewer iterations. In \cite{b10}, the  recurrent neural decoder architecture based on the method of successive relaxation is proposed for improving the error performance and reducing the computational complexity. The authors in \cite{b11} view the decoding as the classification problem and train the weights of the neural network decoder using the generated  dataset
that contains a large number of codewords. It is observed in \cite{b11} that structured codes such as the polar codes are easier to learn than the random codes and the neural networks are difficult to train for long codes.

Despite a variety of schemes on the deep-learning-based demodulation and decoding, they never analyze the feasibility of using the neural network  to demodulate/decode the complex modulations/codes close to the Shannon capacity. In fact,  modulations/codes that are closer to the Shannon capacity generally have the higher demodulation/decoding complexity. In this paper, we use the GAM and the Gaussian codebook to evaluate the neural network demodulator and decoder,  respectively,  in terms of the error performance and the complexity. The GAM can  overcome the asymptotic shaping loss seen in QAM and offers the Shannon capacity approaching performance\cite{b18}.  Gaussian codebooks have often been used to prove direct coding theorems for the Gaussian channel\cite{b12,b13}. We boldly give our conjecture: the neural network demodulator/decoder is a strong candidate approach for demodulating/decoding the complex modulations/codes close to the Shannon capacity owing to the error performance of  the near-ML algorithm and the lower complexity order. The conjecture is based on the following fact/vision: 1) For the low-order GAM and the small Gaussian codebook, the neural network demodulator/decoder obtains the good performance, i.e., the error performance  close to the ML algorithm and the lower complexity order; 2) The rapid development of computing resources makes it possible for the deep neural network to train the high-order modulation and the long codeword.

\section{System Model}
\subsection{Modulation-Demodulation Framework}
\begin{figure}[!t]
\centering
\includegraphics[width=2.7in]{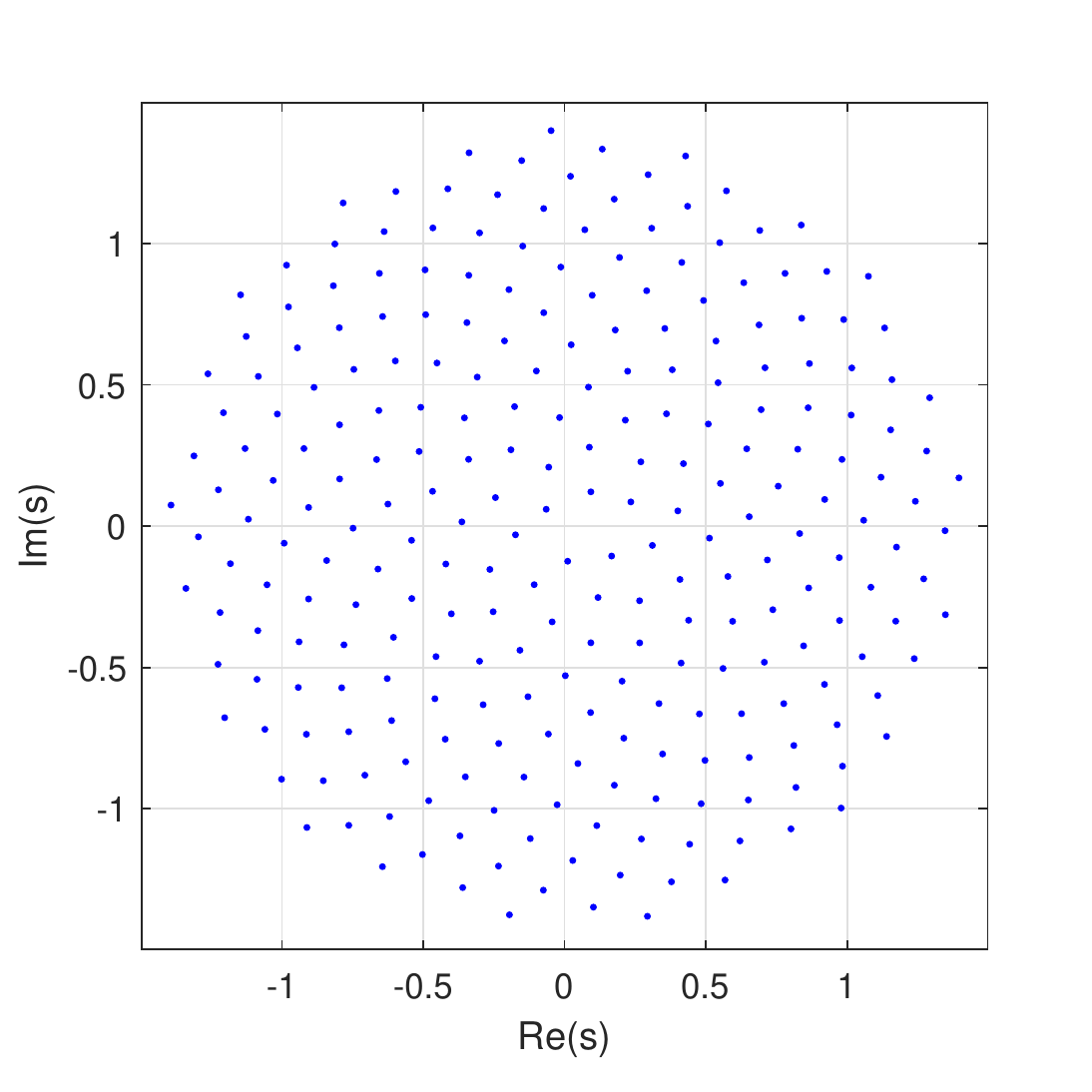}
\caption{Disc-GAM signal constellation, $M=2^8$.}
\label{gam}
\end{figure}

Consider the modulation-demodulation model, in which the transmitted signal is subjected to additive white
Gaussian noise. The received symbol is formulated as
\begin{align}
\bm{y}_1=\bm{x}_1+\bm{\nu}_1
\end{align}
where $\bm{y}_1\in \mathbb{C}^{{n_1}}$ is the received symbol; $\bm{\nu}_1\in \mathbb{C}^{{n_1}}$ is the noise, which follows independent identically distributed zero-mean circularly symmetric complex Gaussian (CSCG) distribution with variance $\sigma_1^{2}$; $\bm{x}_1\in \mathbb{C}^{{n_1}}$ is the transmitted  symbol obtained by modulating the information bit $\bm{b}_1\in\{0, 1\}^{k_1}$. In this paper, we assume that the GAM is used for the transmitter modulator. The $m$th constellation point of GAM can be denoted as
\begin{equation}
s_{m}=a_m{\rm e}^{j2\pi \theta m},~~~~m \in \mathcal{I}_M
\label{r1}
\end{equation}
where $M=2^{k_1}$ is the modulation order; $a_m$ is  the radius of $m$th constellation symbol; $\theta=1-(\sqrt{5}-1)/2$; $2\pi \theta \approx{137.5^\circ}$ is the golden angle in radians; $\mathcal{I}_M=\{1, 2, \dots, M\}$ is defined as a shorthand of the index set. It's noted that $a_{m+1}>a_m$ in golden angle modulation and the later constellation point turns a fixed angle relative to the previous constellation point. Hence, the shape of the constellation is like an increasing spiral. Every constellation point has a unique index and the number of constellation points of GAM can be increased arbitrarily, which  makes the design of GAM constellation more flexible than QAM.  In this paper, we use disc-GAM and its distribution of constellations point is disc-shape (see \cite{b18} in detail). The  amplitude of the $m$th constellation point can be expressed as
\begin{equation}
a_{m}=\sqrt{\frac{2{P_1}m}{M+1}},~~~~m \in \mathcal{I}_M
\label{r2}
\end{equation}
where ${P_1}$ is the average power constraint. An example of disc-GAM is illustrated in Fig. \ref{gam}.

 The receiver gets the $\bm{y}_1$ and demodulates it as $\hat{\bm{b}}_1$ using the demodulator, i.e.,
 \begin{align}
 \hat{\bm{b}}_1=\mathcal{G}_1(\bm{y}_1)
 \end{align}
where $\mathcal{G}_1$ denotes the demodulator at the receiver.

\subsection{Coding-Decoding Framework}

We suppose that the information bit $\bm{b}_2\in \{0,1\}^{k_2}$ is mapped into the code sequence $\bm{x}_2\in\mathbb{R}^{n_2}$, where each element is selected at random independent identically distributed according to $\mathcal{N}(0, P_2)$. Here, $P_2$ is the average power constraint and the generated Gaussian codebook is known by both the transmitter and the receiver.
Then, the code sequence $\bm{x}_2$ is transmitted over a channel that is subjected to additive white Gaussian noise. The received signal can be written as
\begin{align}
\bm{y}_2=\bm{x}_2+\bm{\nu}_2
\end{align}
where $\bm{y}_2$ is the channel output and $\bm{\nu}_2$ is the zero-mean Gaussian noise with variance $\sigma_2^2$, i.e., $\bm{\nu}_2\sim\mathcal{N}(0, \sigma_2^2\bm{I}_{n_2})$.

The receiver gets the $\bm{y}_2$ and decodes it as $\hat{\bm{b}}_2$ using the decoder, i.e.,
\begin{align}
 \hat{\bm{b}}_2=\mathcal{G}_2(\bm{y}_2)
 \end{align}
where $\mathcal{G}_2$ denotes the decoder at the receiver.

\section{Deep-learning-based Demodulator/Decoder}
In this section, we firstly introduce the ML algorithm, which is used for the baseline of evaluating the neural network demodulator/decoder. Then, the neural network demodulator/decoder  is introduced in detail. Essentially, the demodulator and the decoder have the same mathematical properties, i.e., recovering the received signal to bits. For convenience, we use $\bm{y}$ to represent $\bm{y}_1$ and $\bm{y}_2$, and similar operations also are applied to $\bm{x}_1$, $\bm{x}_2$, $\bm{b}_1$, $\bm{b}_2$, $\hat{\bm{b}}_1$, $\hat{\bm{b}}_2$, $k_1$, $k_2$, $n_1$ and $n_2$.

\subsection{ML Demodulator/Decoder}
An optimal and universal demodulation/decoding algorithm is named the ML algorithm, which is formulated as
 \begin{align}
\hat{\bm{b}}=\mathop{\text{argmax}}\limits_{\bm{b}\in\mathcal{B}}~p(\bm{y}|\bm{b})
\end{align}
where $\mathcal{B}$ is the set of all possible cases. As shown in Section II, the $M$-order GAM consists of $2^{k_1}$ possible symbols, in which  $\mathcal{B}$ denotes the set of all possible symbols. Meanwhile, the generated codebook contains $2^{k_2}$ codewords, in which $\mathcal{B}$ denotes the set of all possible codewords. The ML algorithm compared the received noisy signal $\bm{y}$ with each of the set $\mathcal{B}$ and picks the one closest to $\bm{y}$.  Although the ML algorithm is the optimal, obviously, it suffer from the $\mathcal{O}(2^k)$ complexity. Thus, the ML algorithm is  impractical in the communication system especially when the length of the information $\bm{b}$ is large.
\subsection{Neural Network Demodulator/Decoder}
In this paper, we design a simple and universal  full-connected feedforward neural network for demodulation/decoding, which consists of $L$ hidden layers. In a more mathematical way, the neural network demodulator/decoder can be represented as
\begin{align} \label{nnd}
\hat{\bm{b}}&=f(\bm{y};\bm{\theta})\\
&=f^{(L)}\left(f^{(L-1)}\left(f^{(L-2)}\left(\dots\left(f^{(1)}\left(\bm{y}\right)\right)\right)\right)\right)
\end{align}
where $\bm{\theta}$ is the optimal parameters of the neural network and the mapping function of the layer $i$ with  the weight $\bm{w}_i\in \mathbb{R}^{o_i\times c_i}$ and  the bias $\bm{e}_i\in \mathbb{R}^{o_i}$  is denoted as $f^{(i)}: \mathbb{R}^{c_i}\rightarrow \mathbb{R}^{o_i}$ , i.e.,
\begin{align}
f^{(i)}(\bm{d}_i)&=g(\bm{w}_i\bm{d}_i+\bm{e}_i)
\end{align}
where $\bm{d}_i\in \mathbb{R}^{c_i}$ is the input of the $i$th layer and $g(\cdot)$ is the activation function. The number of the neuron of the input layer and the output layer  are depended on the length of the channel output $\bm{y}$ and the information bit $\bm{b}$, respectively. Theoretically, the neural network can approximate any continuous function owing to the nonlinear activation functions when  the number of neurons is large enough.

To obtain the optimal $\bm{\theta}$, the neural network needs to be properly trained.  The training set  contains the received signal sample $\bm{y}$ and the information bit $\bm{b}$, which can be denoted as
\begin{align}
\mathcal{D}=\{(\bm{y}, \bm{b})_j\}_{1}^{J}
\end{align}
where $J$ denotes the number of samples in the training set.  The training of the neural network uses the mean squared error (MSE) as the loss function, which is defined as
\begin{align}
\mathcal{L}=\mathbb{E}(\bm{b}-\tilde{\bm{b}})^2
\end{align}
where $\tilde{\bm{b}}$ is the output of the neural network and $\mathbb{E}(\cdot)$ is the operator denoting the mean. Once the optimal $\bm{\theta}$ is obtained, the neural network demodulator/decoder can used for the demodulation/decoding of the received noisy signal in the testing phase by using \eqref{nnd}.
\section{Performance Analysis}
In this section, we evaluate the performance of the neural demodulator/decoder including the error performance and the complexity. For the convenience of representation, the `NN' and `ML' are used to  represent the neural network demodulator/decoder and the maximum likelihood demodulator/decoder, respectively. Furthermore, the signal to noise ratio (SNR) is measured by the $E_{\text{b}}/N_0$, where $E_{\text{b}}$ and $N_0$ denote the energy of a bit and the power spectral density of the noise, respectively.
\subsection{Parameter Setting}
For the demodulation, the  $4$GAM, $16$GAM, $64$GAM,
and $256$GAM are used for the performance evaluation, i.e., $k_1=2, 4, 6, 8$. The oversampling factor is set as $10$, i.e., a symbol contains $n_1=10$ samples. The transmitted signal is normalized, i.e., the average power $P_1=1$. For the decoding, the codeword with the size $k_2=2, 4, 6, 8$ are used for  performance evaluation. The code rate is $0.5$, i.e., $k_2/n_2=0.5$. The code sequence is normalized, i.e., the average power $P_2=1$.

We use a very large neural network  as the architecture of the neural network demodulator/decoder that consists of four hidden layers and the number of the neuron corresponding to each layer are $1024, 512 ,256, 128$, respectively. The size of the training and the testing for each codeword/symbol index are $2^{18}$ and $2^{17}$, respectively. The batch size is $256$, the ReLu is used for the activation function and we use the Adam as the optimizer.

\begin{figure}[!t]
\centering
\includegraphics[width=2.7in]{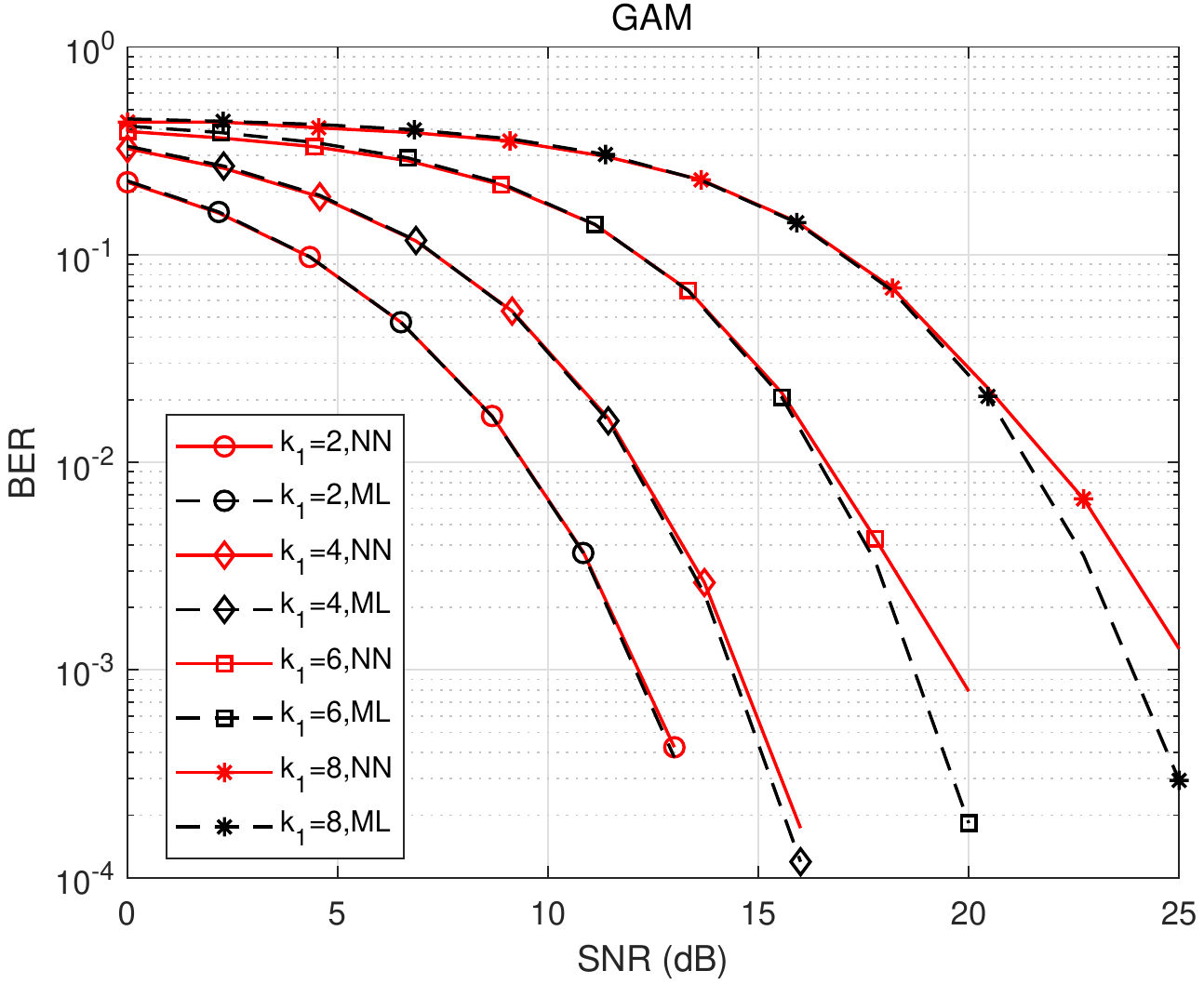}
\caption{The BER  is evaluated for the NN demodulator and the ML demodulator with different modulation order. The results show that the error performance of the NN demodulator is close to
the ML demodulator, especially for low-order modulation.}
\label{snr}
\end{figure}

\begin{figure}[tbp]
  \centering
  \includegraphics[width=2.7in]{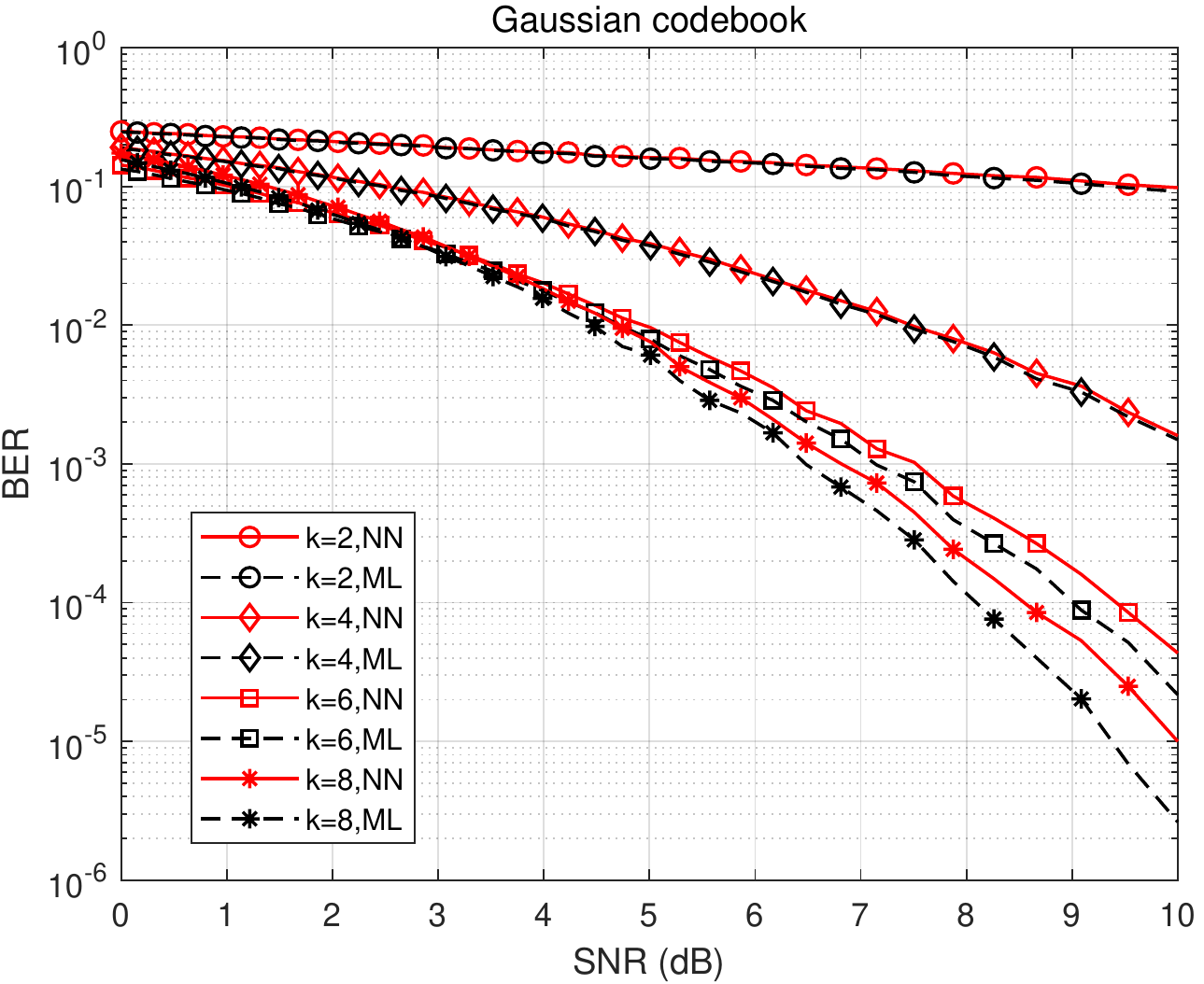}\\
  \caption{The BER  is evaluated for the NN decoder and the ML decoder with different length of the information bit. The results indicate that the NN decoder can achieve near-ML BER and  the shorter length of the information bit, the closer is the gap between NN and ML performance.}  \label{BER}
\end{figure}
\begin{figure}[tbp]
  \centering
  \includegraphics[width=2.7in]{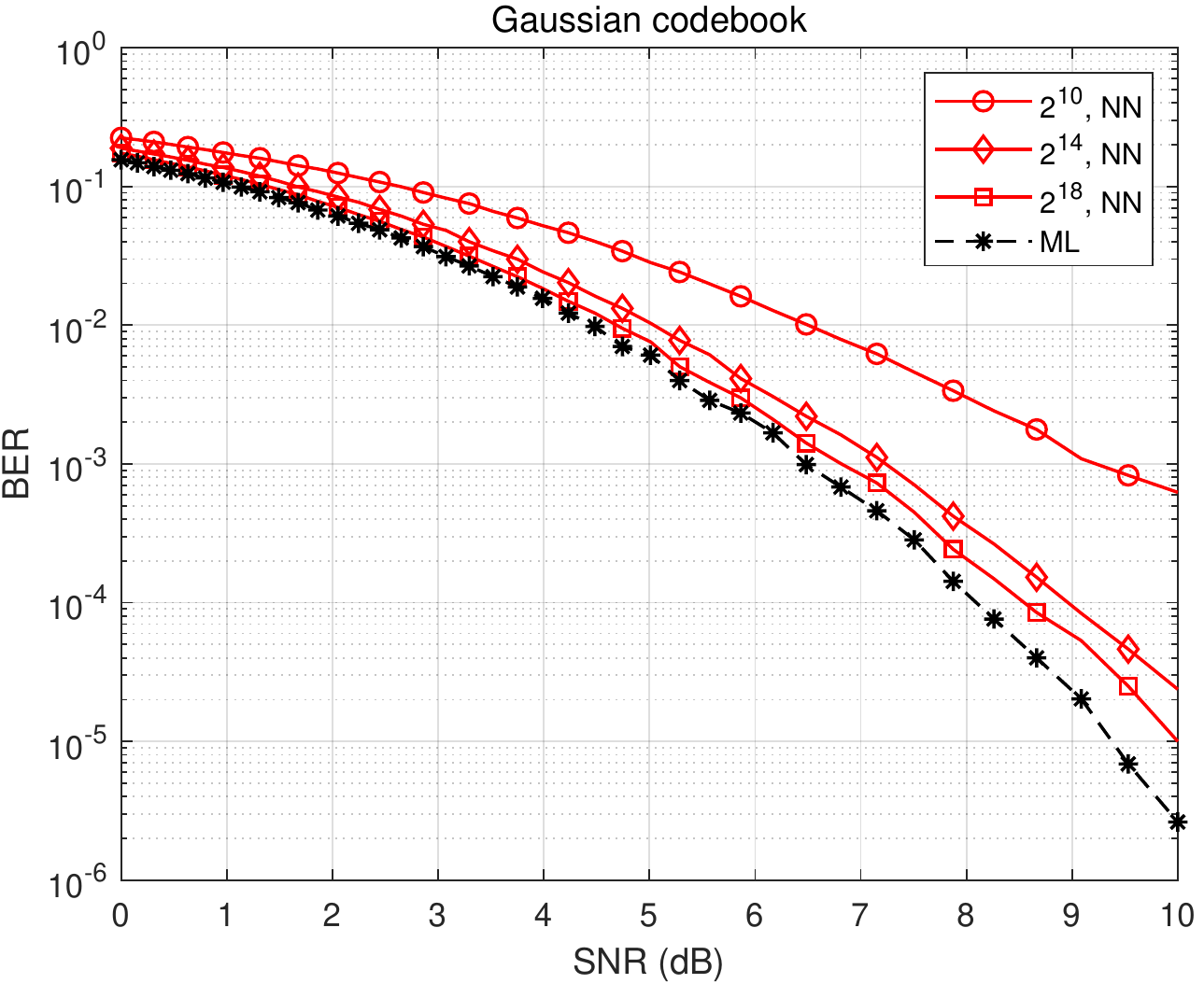}\\
  \caption{The BER  is evaluated for the NN decoder and the ML decoder with different size of training samples when $k_2=8$. The results show that the larger the number of training samples,
the closer is the gap between NN and ML performance.}  \label{BER_TR}
\end{figure}
\subsection{Numerical Results}
\noindent\emph{\textbf{Observation 1:} The neural network demodulator achieves the error performance close to the ML demodulator when the modulation order is low. (cf. Fig. \ref{snr})}

In Fig. \ref{snr}, the BER performance of the NN demodulator and the ML demodulator are evaluated for the  $4$GAM, $16$GAM, $64$GAM,
and $256$GAM. Obviously, the performance of the NN demodulator is close to
the ML demodulator, especially for low-order modulation. This is because that the set $\mathcal{B}$ for the low-order modulation contains fewer possible symbols than the high-order modulation. As a result, the mapping function of the GAM for the  low-order modulation can be learned more easily by the neural network.

\noindent\emph{\textbf{Observation 2:} The neural network demodulator has a lower complexity order compared with the ML demodulator. (cf. Table \ref{tab1})}

In Table \ref{tab1}, the computational complexity of the NN demodulator and the ML demodulator are evaluated for the  $4$GAM, $16$GAM, $64$GAM,
and $256$GAM. We can see that the complexity order of the  NN demodulator is lower than the ML demodulator.
Furthermore, Table \ref{tab1} also explicitly provides numerical results of time consumption of each implementation.
It is observed that the computational time of the ML demodulator increases rapidly with $k_1$ while that of the NN demodulator does not grow so fast. The trend of time computation is consistent with the complexity evaluation. Thus, we conclude that the NN demodulator has a lower complexity order compared with the ML demodulator.

\noindent\emph{\textbf{Observation 3:} The neural network decoder achieves the error performance close to the ML decoder when  the codeword is short. (cf. Figs. \ref{BER} and \ref{BER_TR} )}

In Fig. \ref{BER}, we plot the BER curve of the NN  and the ML decoder  for $k_2=2, 4, 6, 8$. It is observed that the NN decoder can achieve the near-ML performance especially when  the codeword is short. Furthermore, we give the BER of the NN and ML decoder when $k_2=8$ and the size of the training samples is $2^{10}, 2^{14}, 2^{18}$. It can be seen that the large training size can improve the error performance of the NN decoder and the near-ML performance is  achieved when the training size is $2^{18}$. Thus, we conclude that the NN decoder trained with a sufficiently large number of samples can achieve the near-ML performance for the given large-dimensional neural network.

\noindent\emph{\textbf{Observation 4:} The neural network decoder has a lower complexity order compared with the ML decoder. (cf. Table \ref{tab2})}

Table \ref{tab2} illustrates the computational complexity of the decoding for the NN and ML decoders with different codeword sizes. Obviously, the computational complexity order of the NN decoder is significantly less than that of the ML decoder when decoding the same codeword. Furthermore, we also see that as the length of the codeword grows, the time consumption of the ML decoder grows much faster than the NN decoder. Thus, we can conclude that the NN decoder has a lower complexity order compared with the ML decoder.

\begin{table*}[tbp]
\setlength{\abovecaptionskip}{-0.06cm}
\setlength{\belowcaptionskip}{-10cm}
  \centering
  \fontsize{7}{8}\selectfont
  \caption{Computational Complexity of Different Demodulation Methods}
  \label{3}
    \begin{tabular}{c|c|ccccccccc}
    \toprule
    \multirow{3}{*}{Demodulator}&\multirow{3}{*}{Complexity order}&\multicolumn{4}{c}{1 implementation (Sec.)}\\
     \cmidrule(r){3-6}
    &&$k_1=2$&$k_1=4$&$k_1=6$&$k_1=8$\cr
    \rowcolor{red!10}
    \midrule
    NN &$\mathcal{O}\left(2n_1o_1+2o_Lk_1+\sum_{i=2}^{L}2c_io_i\right)$&$0.13850$ & $0.16084$ & $0.17248$ & $0.19001$\cr
    \rowcolor{blue!10}
    \midrule
    ML &$\mathcal{O}\left(2^{k_1}\right)$&$0.00770$ & $0.02611$ & $0.09803$ & $0.41249$\cr
    \bottomrule
    \end{tabular}
    \label{tab1}
\end{table*}

\begin{table*}[tbp]
\setlength{\abovecaptionskip}{-0.06cm}
\setlength{\belowcaptionskip}{-10cm}
  \centering
  \fontsize{7}{8}\selectfont
  \caption{Computational Complexity of Different Decoding Methods}
  \label{3}
    \begin{tabular}{c|c|ccccccccc}
    \toprule
    \multirow{3}{*}{Decoder}&\multirow{3}{*}{Complexity order}&\multicolumn{4}{c}{1 implementation (Sec.)}\\
     \cmidrule(r){3-6}
    &&$k_2=2$&$k_2=4$&$k_2=6$&$k_2=8$\cr
    \rowcolor{red!10}
    \midrule
    NN &$\mathcal{O}\left(2n_2o_1+2o_Lk_2+\sum_{i=2}^{L}2c_io_i\right)$&$0.11401$ & $0.11862$ & $0.12117$ & $0.12696$\cr
    \rowcolor{blue!10}
    \midrule
    ML &$\mathcal{O}\left(2^{k_2}\right)$&$0.01703$& $0.09540$ & $0.56546$ & $3.30310$\cr
    \bottomrule
    \end{tabular}
    \label{tab2}
\end{table*}

\noindent\emph{\textbf{Conjecture:} The neural network demodulator/decoder is a strong candidate approach for demodulating/decoding the complex modulations/codes close to the Shannon capacity owing to the error performance of the near-ML algorithm and the lower complexity order.}

In the observation $1$, $2$, $3$, and $4$, we can conclude that the NN demodulator/decoder has much lower complexity order while achieving error performance close to the ML algorithm in the case of  the low-order modulation and the short codeword. Limited by the current training resources, we cannot evaluate the performance of the high-order modulation and the long codeword such as $k_1=10000$ and $k_2=10000$. But, with the development of offline resources such as the graphical processing units (GPUs), the performance evaluation  of the high-order modulation and the long codeword will be implemented. In fact, the Shannon capacity can be obtained when the constellation point or the code sequence is Gaussian distributed and the modulation order or the length of the codeword is infinite. Here, based on the results of the low-order GAM and the small Gaussian codebook, we boldly give our conjecture: the neural network demodulator/decoder is a strong candidate approach  for demodulating/decoding the complex modulations/codes close to the Shannon capacity owing to the error performance of  the near-ML algorithm and the lower complexity order.

\section{Conclusion}
In this paper, the feasibility of using the neural network to demodulate/decode the complex modulations/codes close to the Shannon capacity is analyzed. We evaluate the performance of the neural network demodulator/decoder in the case of the low-order GAM and the small Gaussian codebook. The results show that the NN demodulator/decoder has much lower complexity order while achieving error performance close to the ML algorithm. Based on the promising results of  the low-order GAM and the small Gaussian codebook, we also give our conjecture: the neural network demodulator/decoder is a strong candidate approach for demodulating/decoding the complex modulations/codes close to the Shannon capacity owing to the error performance of  the near-ML algorithm and the lower complexity order.


\begin{thebibliography}{1}
\bibitem{b1} C. E. Shannon, ``A mathematical theory of communication,"
\emph{Bell System Technical Journal}, 1948.
\bibitem{b14} G. D. Forney and G. Ungerboeck, ``Modulation and coding for linear Gaussian channels," \emph{IEEE Trans. Inf. Theory,} vol. IT-44, no. 6, pp. 2384-2415, Oct. 1998.
\bibitem{b15} G. Forney, ``Trellis shaping," \emph{IEEE Trans. Inf. Theory,} vol. 38, no. 2, pp. 281-300, 1992.
\bibitem{b16} W. Betts, A. Calderbank, and R. Laroia, ``Performance
of nonuniform constellations on the gaussian channel," \emph{IEEE Trans. Inf. Theory,} vol. 40, no. 5, pp. 1633-1638, 1994.
\bibitem{b17} D. Divsalar, M. Simon, and J. Yuen, ``Trellis coding
with asymmetric modulations," \emph{IEEE Trans. Commun.,} vol. 35, no. 2, pp. 130-141, 1987.
\bibitem{b18} P. Larsson, ``Golden angle modulation: Geometric- and probabilistic-shaping," 2017, [online] Available: https://arxiv.org/abs/1708.07321.
\bibitem{b2} R. Gallager, ``Low-density parity-check codes," \emph{IEEE Trans. Inf. Theory,} vol. 8, no. 1, pp. 21-28, 1962.
\bibitem{b3} C. Berrou, A. Glavieux, and P. Thitimajshima, ``Near shannon
limit error-correcting coding and decoding: Turbo-codes. 1,"
in \emph{Proc. IEEE Int. Conf. Commun. (ICC),}
vol. 2, May 1993, pp. 1064-1070.
\bibitem{b4} E. Arikan, ``Channel polarization: A method for constructing
capacity-achieving codes for symmetric binary-input memoryless channels," \emph{IEEE Trans. Inf. Theory,} vol. 55, no. 7, pp. 3051-3073, 2009.
\bibitem{b19} H.-G. Yeh and H. Seo, ``Low complexity demodulator
for M-ary QAM," in \emph{Proc. Wireless Telecommun. Symp.,} 2007, pp. 1-6.
\bibitem{b20} A. S. Mohammad, N. Reddy, F. James, and C. Beard,
``Demodulation of faded wireless signals using deep
convolutional neural networks," in \emph{Proc. IEEE 8th Annu. Comput. Commun. Workshop Conf.,} 2018, pp. 969-975.
\bibitem{b21} L. Fang and L. Wu, ``Deep learning detection method for
signal demodulation in short range multipath channel,"
in \emph{IEEE 2nd Int. Conf. Opto Electron. Inf. Process. (ICOIP),} 2017, pp. 16-20.
\bibitem{b22} T. Wu, ``CNN and RNN-based deep learning methods for digital signal demodulation," in \emph{Proc. Int. Conf. Image Video Signal Process. (IVSP),} 2019, pp. 122-127.
\bibitem{b23} S. Zheng, X. Zhou, S. Chen, P. Qi, and X. Yang,
``Demodnet: Learning soft demodulation from hard information using convolutional neural network," arXiv: Signal Processing, 2020.
\bibitem{b5} J. Pearl, Probabilistic Reasoning in Intelligent Systems: Networks of Plausible Inference, \emph{CA, San Mateo:Morgan Kaufmann,} 1988.
\bibitem{b6} C. Berrou, A. Glavieux, and P. Thitimajshima, ``Near Shannon limit error-correcting coding and decoding: Turbo-codes," in \emph{Proc. Int. Conf. Commun. (ICC),} May 1993, pp. 1064-1070.
\bibitem{b7} M. P. C. Fossorier and S. Lin, ``Soft-decision decoding of linear block codes based on ordered statistics," \emph{IEEE Trans. Inf. Theory,} vol. 41, no. 5, pp. 1379-1396, 1995.
\bibitem{b8} K. R. Duffy, J. Li, and M. Medard, ``Capacity-achieving guessing random additive noise decoding," \emph{IEEE Trans. Inf. Theory,} vol. 65, no. 7, pp. 4023-4040, 2019.
\bibitem{b9} E. Nachmani, Y. Be'ery, and D. Burshtein, ``Learning to decode linear codes using deep learning," in \emph{Proc. IEEE Annu. Allerton Conf. Commun. Control Comput. (Allerton),}, 2016, pp. 341-346.
\bibitem{b10} E. Nachmani, E. Marciano, L. Lugosch, W. J. Gross, D. Burshtein, and Y. Be'ery, ``Deep Learning Methods for Improved Decoding of Linear Codes," \emph{IEEE J. Sel. Topics Signal Process.,} vol. 12, no. 1, pp. 119-131, Feb. 2018.
\bibitem{b11} T. Gruber, S. Cammerer, J. Hoydis, and S. t. Brink, ``On deep learning-based channel decoding," in \emph{Proc. Annual Conf. on Inf. Sciences and Systems (CISS),} 2017, pp. 1-6.
\bibitem{b12} T. M. Cover and J. A. Thomas, Elements of information theory. John Wiley and Sons, New York, 1991.
\bibitem{b13} R. G. Gallager, Information theory and reliable communication. John Wiley and Sons, New York, 1968.
\end{thebibliography}
\end{document}